\documentclass[aps,prb,amssymb,twocolumn,showpacs]{revtex4}
\usepackage{tabularx,graphicx,float}
\newcommand{\bk}{\vec{\bf k}}
\newcommand{\bq}{\vec{\bf q}}
\newcommand{\ek}{\epsilon_{\bk}}
\newcommand{\ekq}{\epsilon_{\bk+\bq}}
\newcommand{\ef}{\epsilon_{\rm F}}
\newcommand{\cx}{\cos ( k_x a )}
\newcommand{\cy}{\cos ( k_y a )}
\newcommand{\sx}{\sin ( k_x a )}
\newcommand{\sy}{\sin ( k_y a )}

\newcommand{\susc}{\Pi ( \bq , \omega )}
\newcommand{\vf}{v_{\rm F}}

\begin{document}
\title{Self-energy corrections to anisotropic Fermi surfaces.}
\author{R. Rold\'{a}n$^1$, M.P. L\'{o}pez-Sancho$^1$, F. Guinea$^1$, and
  S.-W. Tsai$^2$}
\affiliation{{$^1$Instituto de Ciencia de Materiales de Madrid,
CSIC, Cantoblanco, E-28049 Madrid, Spain.\\
$^2$Department of Physics, University of California, Riverside, CA 92521, USA}
}
\date{\today}
\begin{abstract}

The electron-electron interactions affect the low-energy
excitations of an electronic system and induce deformations of the
Fermi surface. These effects are especially important in
anisotropic materials with strong correlations, such as copper
oxides superconductors or ruthenates. Here we analyze the
deformations produced by electronic correlations in the Fermi
surface of anisotropic two-dimensional systems, treating the
regular and singular regions of the Fermi surface on the same
footing. Simple analytical expressions are obtained for the
corrections, based on local features of the Fermi surface. It is
shown that, even for weak local interactions, the behavior of the
self-energy is non trivial, showing a momentum dependence and a
self-consistent interplay with the Fermi surface topology. Results
are compared to experimental observations and to other theoretical
results.

\end{abstract}
\maketitle

\section{Introduction}

Anisotropic materials present different physics at different
energy scales, and their behavior or response to external probes
is difficult to interpret. A big amount of experimental works have
made possible to study the puzzling electronic properties of many
anisotropic materials which, in general, present potential
technological applications. More theoretical effort is needed in
order to understand the  detailed experimental data which reveal
an unconventional behavior. In conventional metals, the
excitations that govern their low-temperature physics present well
defined momenta lying at the three-dimensional Fermi surfaces. In
the anisotropic materials, as layered transition metal oxides,
unusual electronic properties appear and, under certain
conditions, changes of the effective dimensionality occur. The
electronic  interaction effects are enhanced as the dimensionality
decreases and can change the fundamental properties of the
material\cite{Valla02}. Therefore, due to both the anisotropy and
the periodicity along the axis perpendicular to the planes,
specific collective excitations appear absent in two-dimensional
(2D) and three-dimensional (3D) electron
gases\cite{Arri00,Bierm01}.

The high-temperature cuprate superconductors are among the most
studied layered transition metals oxides, treated as 2D systems in
many approaches, due to its strong anisotropy. In the hole doped
cuprates the FS topology changes with doping from hole-like to
electron-like \cite{Yosh,Kami}. Recently a change in the sign of
the Hall coefficient has been reported for heavily overdoped
LaSrCuO$_4$ \cite{Tsuk}. The evolution of the FS in electron-doped
copper oxide superconductors with doping has been reported by
ARPES experiments to change from electron-pocket centered at the
$(\pi, 0)$ point of the Brillouin zone at low doping to a
hole-like FS centered at $(\pi,\pi)$ at higher
doping\cite{Armit02}.

Other transition metal oxides as cobaltates or ruthenates are
multiorbital systems and their FS present a complex topology with
different sheets derived from the different bands at the Fermi
energy. A correlated 2D material particularly interesting is the
Sr$_2$RuO$_4$, a ruthenate considered a  model Fermi liquid (FL)
system with important electronic correlations which have to be
taken into account when interpreting photemission
spectra\cite{Ingle05} to obtain a clear picture of the electronic
properties, especially in the vicinity of the Fermi energy. In
Sr$_2$RuO$_4$ the FS separates into three sheets $\alpha$,
$\beta$, and $\gamma$, coming from the $d_{xz}$, $d_{yz}$ and
$d_{xy}$ orbitals.

In the study of the electron-electron
interactions in anisotropic metallic systems an open question is
the deformation of the Fermi surface induced by these
interactions. The Fermi surface is one of the key features needed
to understand the physical properties of a material and its shape
provides important information. Recent improvements in
experimental resolution have led to high precision measurements of
the Fermi surface, and also to the determination of the many-body
effects in the spectral function, as reported by angle resolved
photoemission spectroscopy (ARPES) experiments \cite{DHS03}.
However the interpretation of  the data obtained by different
experimental techniques in anisotropic strongly correlated systems
remains a complex task\cite{Zhou05}.

The Fermi surface  depends on the self-energy corrections to the
quasiparticle energies, which, in turn, depend on the shape of the
Fermi surface. Hence, there is an interplay between the
self-energy corrections and the Fermi surface topology. For weak
local interactions, the leading corrections to the Fermi surface
(FS) arise from second order diagrams. The self energy, within
this approximation, can show a significant momentum dependence
when the initial FS is anisotropic and lies near hot spots, where
the quasiparticles are strongly scattered\cite{RR06}. This
simultaneous calculation of the FS and the second order
self-energy corrections is a formidable task. However, the
knowledge of the exact shape of the FS of a material is very
important since it may affect the transport properties as well as
the collective behavior, and have a valuable information from the
point of view of theory in orther to find the appropriate model to
study the system. Many approaches have been used to study  this
problem like mean-field\cite{VV01,V05}, pertubation
theory\cite{ViRu90}, bosonization methods\cite{AHCN94,FSLut99}, or
perturbative Renormalization Group
calculations\cite{AlIoMi95,Metz03,Kop03,FCF05,DD03}, and the cellular
dynamical mean-field theory (CDMFT), an extension of Dynamical
Mean Field Theory\cite{Kotl05}, and many others. In spite of the
great theoretical effort done in the last years, there is a need
to develop alternative new methods in order to understand the
origin of the electronic properties in  materials with strong
correlations.

In this work, we calculate perturbative corrections and use
Renormalization Group arguments\cite{S94,MCC98} in order to study
analytically the qualitative corrections to the shape of the FS
induced by the electron-electron interaction. This method allows
us to classify the different features of the FS from the
dependence of the self-energy corrections on the value of the high
energy cutoff, $\Lambda$, defined at the beginning of the
Renormalization process. As it will be shown later, one can also
analyze the effects of variations in the Fermi velocity and the
curvature of the non interacting FS on the self-energy
corrections. The calculations do not depend on the microscopic
model which gives rise to a particular Fermi surface, so that it
can be useful in different situations. For concreteness we will
consider the $t-t'$ Hubbard model to study two dimensional Fermi
surfaces of cuprates and an extension of it to study the case of
Sr$_2$RuO$_4$. The paper is organized as follows. We define the
model in Section II and describe the way the self-energy corrections
are calculated. In Section III we present a detailed calculation of the
changes expected for a regular FS, as well as for a FS showing
singular points like Van Hove singularities, nesting or inflexion
points. We compare with results from ARPES experiments on
anisotropic materials, mainly cuprate superconductors and
Sr$_2$RuO$_4$. In the last section we highlight the most relevant
aspects of our calculation, and some conclusions are presented.

\section{The method}
The method of calculation of the self-energy corrections does not
depend on the microscopic model used to obtain the electronic
structure and the FS. In our scheme  simple analytical expressions
of the effects induced by the interactions are deduced from local
features of the Fermi surface, and we are able to treat, on the
same footing, the regular and singular regions of the FS.
Therefore the method is particularly useful in correlated
anisotropic materials which present exotic properties and deviate
from band structure calculations. The importance of considering
correlation effects when interpreting experimental data is already
known and recently a great effort has been made in order to
evaluate the self-energy from ARPES spectra\cite{Kordy05}. The
evaluation of many body effects in these complex materials is far
from trivial since the electron scattering presents a dependence
on momentum and energy. We limit the study to the weak coupling
regime, considering weak local interactions, consistent with the
Hubbard model.

\subsection{The model.}
We consider the $t-t'$ Hubbard model which is the simplest
theoretical model which allows us to study different correlated
materials and describes the shape of the FS observed by ARPES in
different materials as cuprates (see ref.\cite{DHS03} and
references therein). Depending on the ratio $t'/t$ and on the
band-filling, different phases and stabilities appear, as found in
early mean-field and Quantum Monte Carlo (QMC) studies of the
model\cite{Lin87}. By changing the parameters a rich phase
diagram, including antiferromagnetic, ferromagnetic and
superconducting phases, has been found for the 2D $t-t'$ Hubbard
model which describes many physical features of copper oxides and
of Sr$_2$RuO$_4$\cite{Katan03}. For the cuprates, the most studied
model is the Hubbard model on a square lattice and considering an
effective single band. The Hamiltonian of the $t-t'$ Hubbard model
is:
\begin{equation}
{\cal H}\!=\sum_{s; i,j }\!t_{ij} c^\dag_{s,i} c_{s,j} + U\!
\sum_i\! n_{i \uparrow} n_{i \downarrow} \label{hamil}
\end{equation} where $ c_{s,i} (c^\dag_{s,i})$ are destruction
(creation) operators for electrons of spin $s$ on site $i$,
$n_{i,s}=c^\dag_{s,i} c_{s,i}$ is the number operator, $U$ is the
on-site repulsion, and $t_{ij}=t$ are the nearest and $t_{ij}=t'$
the next-nearest neighbors hopping amplitudes, respectively. The
Fermi surfaces of the non interacting systems are defined by:
\begin{equation}
\ef = \varepsilon (\bk) = 2 t \left[ \cx + \cy \right] + 4 t' \cx
\cy \label{dispersion} \end{equation} where $a$ is the lattice
constant.

\begin{figure}[h]
\centerline{\includegraphics*[bb=150bp 40bp 600bp
460bp,clip,scale=0.25]
{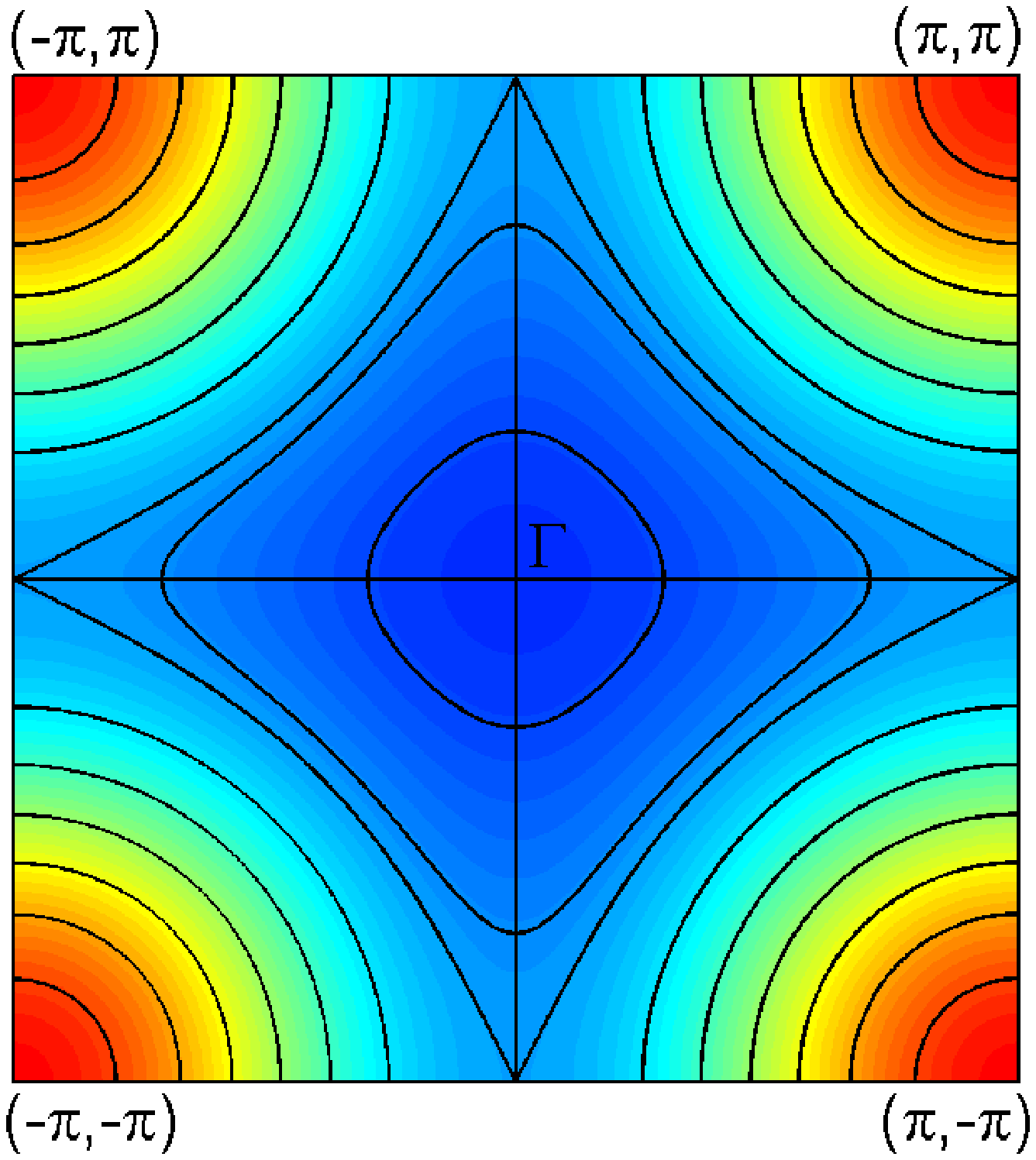}
\includegraphics*[
bb=150bp 40bp 600bp 460bp,clip,scale=0.25]{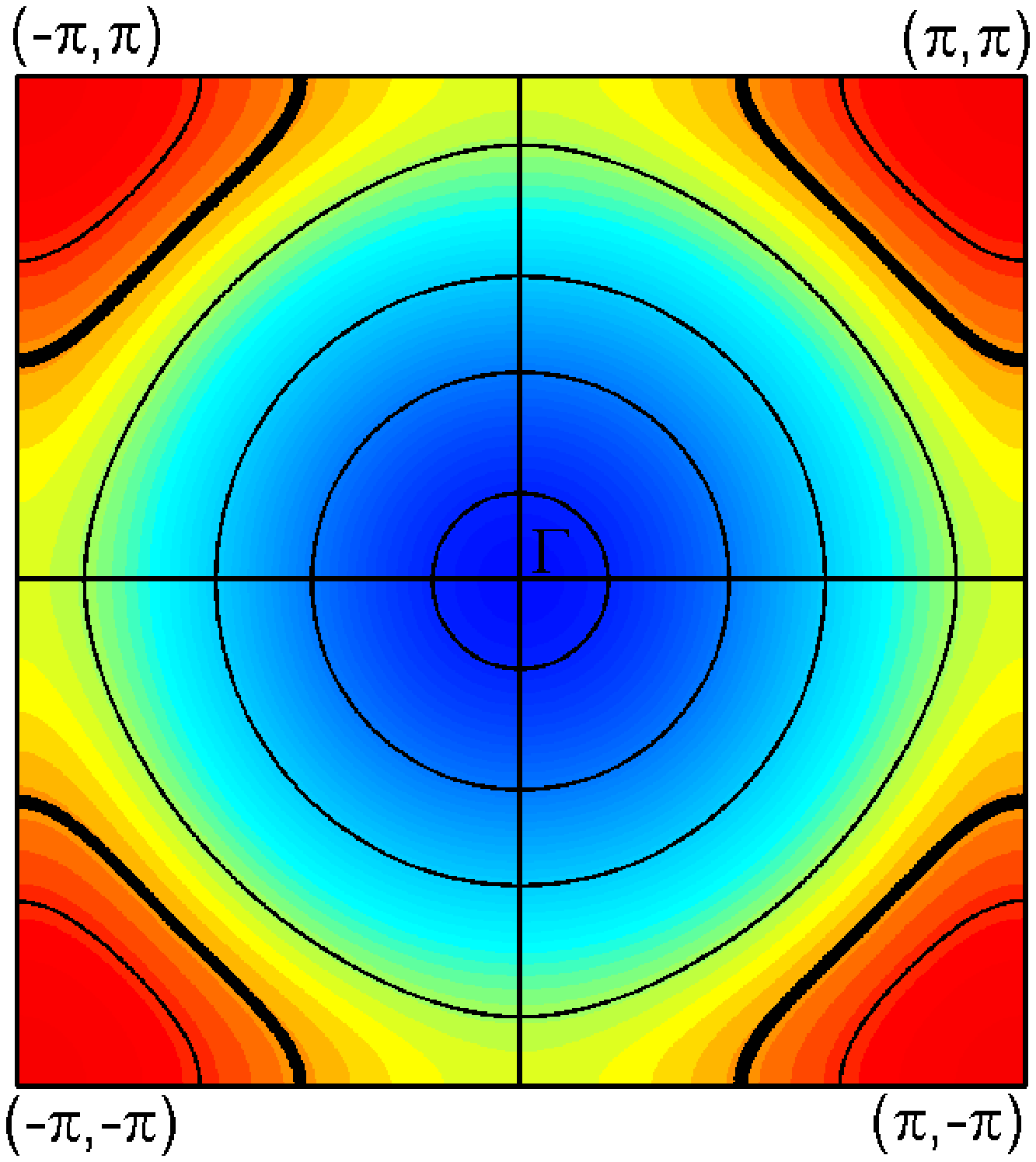}}
\caption{(Color online) Qualitative picture of the evolution of
the FS with filling from almost isotropic to convex, going through
a FS exhibiting inflexion points, and one with van Hove
singularities (left panel, $t'=-0.3t$). A region with almost
perfect nesting is shown in the right panel ($t'=0.3t$).}
\label{FS-Desnudas}
\end{figure}

Assuming that $t < 0 , t' > 0$ and $| 2 t' | < | t |$, the Fermi
surface is convex for $- 2 t + 4 t' \le \ef \le \epsilon_0 = - 8
t' + 16 t'^3 / t^2$. For $ - 8 t' + 16 t'^3 / t^2 \le \ef \le - 4
t'$  the Fermi surface shows eight inflexion points, which begin
at $k_x  = k_y  = k_0 = a^{-1} \cos^{-1} ( - 2 t' / t )$ and move
symmetrically around the $( \pm 1 , \pm 1 )$ directions, towards
the center of the edges of the square Brillouin zone, $( 0 , \pm
\pi ) , ( \pm \pi , 0 )$. For $\ef = 4 t'$ the Fermi surface
passes through the saddle points (Van Hove singularities) located
at these special points of the Brillouin zone. For $4 t' < \ef \le
- 4 t$, the Fermi surface is convex and hole like, centered at the
corners of the Brillouin Zone, $ ( \pm \pi , \pm \pi )$. In
Fig.[\ref{FS-Desnudas}] the variation of the FS shapes with doping
is qualitatively shown.  When  only nearest-neighbor hopping is
considered, $t'=0$, the model has particle hole symmetry, and the
Fermi surface shows perfect nesting for $\ef = 0$.  FS shapes
similar to these shown in Fig.[\ref{FS-Desnudas}] have been
experimentally observed by ARPES on different cuprate samples, at
different doping levels.

\subsection{Self-energy analysis.}

\begin{figure}[b]
\resizebox{6cm}{!}{\includegraphics[]{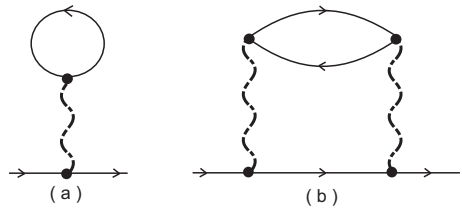}}
\caption{Low order self-energy diagrams. Left: Hartree diagram.
Right: two loop correction.} \label{selfenergy}
\end{figure}

We will analyze the interplay between the  electron-electron
interactions and the FS topology in the weak-coupling regime.
The corrections to the non-interacting Fermi surface are given by
the real part of the self-energy. For each filling $n$, a Fermi
surface is defined. The electron-electron interaction leads to a
self-energy

\begin{equation}
 \Sigma ( \bk , \omega ) = {\rm Re} \Sigma ( \bk , \omega )+ i {\rm Im} \Sigma ( \bk , \omega )
\label{selfen}
\end{equation}

which modifies the bare one-particle propagator
$G_0(\bk,\omega)^{-1}\!=\!\omega\!-\!\ek\!+\!i\delta \rm
\,sgn\,\ek$ (where $\delta\rightarrow 0^+$) to

\begin{equation}
G ( {\bf \vec{k}} , \omega ) = \frac{1}{\omega-[ \varepsilon
(\bk)-\ef]- \Sigma ( \bk , \omega )} \label{green}
\end{equation}
and the FS of the interacting system is given by the $\omega=0$
solution of the equation
\begin{equation}
\ef-\varepsilon (\bk)-{\rm Re} \Sigma ( \bk , \omega )=0
\label{dFS}
\end{equation}
where ${\rm Re} \Sigma ( \bk , \omega )$ is the real part of the
self-energy. The diagrams that renormalize the one-particle Green
function up to second order in perturbation theory are depicted in
Figure[\ref{selfenergy}]. The Hartree diagram, shown at the left
of the figure, gives a contribution which is independent of
momentum and energy, hence it cannot deform the FS. The two-loop
diagram (right of Fig.[\ref{selfenergy}]), modifies the FS through
its $\bk$ dependence and, in addition, it changes the
quasiparticle-weight through its $\omega$ dependence.

As explained above there are many  possible shapes of the FS which
fit the experimental results from copper-oxide or ruthenate
samples. The conventional perturbation theory fails on describing
FS for which logarithmic divergences in the density of states
(DOS) appear at certain values of the parameters of
Eq.(\ref{hamil}). Then we proceed to calculate the self-energy by
adopting a Renormalization Group strategy\cite{S94,MCC98}. It is
assumed that the effect of the high energy electron-hole pairs on
the quasiparticles near the Fermi surface have been integrated
out, leading to a renormalization of the parameters $t , t'$ and
$U$ of the hamiltonian. The possibility that other couplings are
generated in the system is not allowed. Thus, the Hamiltonian,
Eq.(\ref{hamil}), describes low temperature processes below a high
energy cutoff, $\Lambda \ll t , t'$. For consistency, we consider
the Hubbard interaction $U \lesssim \Lambda$,  as well below the
energy cutoff. Therefore the corrections to the quasiparticle
energies are determined as a function of $\Lambda$, which defines
an energy scale about the Fermi line which will contain the modes
we are interested on (low energy excitations with
$|\ek|<\Lambda$), separated from the high energy excitations (with
$|\ek|>\Lambda$) which will be integrated out. We have to notice
that we are restricting ourselves to a momentum-independent
coupling $U$ which corresponds to a local interaction in the real
space. During the process we assume that the Fermi surface of the
interacting system exists, and that this FS, dressed by the
corrections due to the interactions, has the same topology as that
of the non-interacting system.

The two-loop self-energy shown in Fig.[\ref{selfenergy}](b) can be
computed from

\begin{equation}\label{selfenergy-full}
 i\Sigma_2(\bk,\omega)=\frac{1}{(2\pi)^3}{\int}\!d\omega^{\prime}\!\!\int\!
 d^2\!q \rm G_0(\bk-\bq,\omega-\omega^{\prime})\Pi( \bq , \omega^{\prime} )
\end{equation}

where the one-loop particle-hole polarizability, in terms of the
one-particle propagator, reads

\begin{widetext}
\begin{equation}\label{polarizability}
i\susc=\frac{U^2}{(2\pi)^3}\int\!
d\omega^{\prime}\!\!\int\!{d^2\!k\,\Theta(\Lambda-|\ek|)\Theta(\Lambda-|\ekq|)G_0(\bk,\omega^{\prime})G_0(\bk+\bq,\omega+\omega^{\prime})}
\end{equation}
\end{widetext}

The cutoff in energies $\Lambda$ is used to implement the RG
scheme\cite{S94}: the virtual states in the loop of the diagram
shown in Fig.[\ref{selfenergy}](b) have to be kept in the energy
range determined by the cutoff.

\section{Results}

After we compute the self-energy as explained above, we will
analyze the deformation induced in the Fermi surface shape. We are
interested in anisotropic two-dimensional FS similar to those
measured for the cuprates which present regions with different
scattering rates. For clarity, we consider first Fermi surfaces
with hot spots, which are regions where the scattering of the
quasiparticles is strongly enhanced. At these points, the
scattering could be singular giving divergences of the
susceptibility. Separately we address the deformations of regular
FS, curved surfaces which do not have singularities. These
surfaces present a scattering rate relatively weak. We will show
that once the Fermi velocity and curvature of the non-interacting
FS are known, we can evaluate the corrections to the FS shape.
Even considering a weak local interaction, without momentum
dependence, the effects are strongly dependent of the location at
the FS.

\subsection{Self-energy corrections to the Fermi surface near hot spots.}

In anisotropic materials, the FS can present regions or special
points which are called hot spots where the quasiparticles become
strongly scattered and their behavior deviates from the
conventional Landau Fermi liquid. The FS of layered transition
oxides, as have been shown above, changes with doping adopting
different shapes which lie close to Van Hove singularities, or
present nested flat regions or inflexion points combined with
regular sectors.

The effects of the Hubbard interaction have been studied when the
Fermi surface presents hot spots both, near the perfect
nesting\cite{ZYD97,AD97,F03c} or near a van Hove
singularity\cite{LB87,F87,D87,S87,LMP87,MG89,Netal92,GGV96}. The
curvature of the FS has important implications in the properties
of the system and the inflexion points, which separate regions
where the curvature has opposite signs, may induce anomalous
effects\cite{GGV97b,FG02}. The crucial role that the FS geometry
plays on the unusual physics of 2D systems, makes it desirable a
deeper insight in the interplay between it and measurable
parameters. The functional dependence of the self-energy on the
cutoff is different in the vicinity of the hot spots than in
regular zones of the FS. Near the hot spots to be considered here,
the dispersion relation satisfies, near the Fermi level:

\begin{equation}
\ek \approx \left\{\begin{array}{cc}
  \pm \frac{k_x^2}{m_x} \mp \frac{k_y^2}{m_y} & \textrm{Van Hove} \\
  v_Fk_{\perp} & \textrm{nesting} \\
\end{array}\right.
\end{equation}

where $\ek=\varepsilon (\bk)-\ef$, $k_{\perp}$ is the momentum
perpendicular to the FS relative to $k_F$,
$k_{\perp}=(\bk-\bk_F)_{\perp}$, $v_F$ is the Fermi velocity at
any particular point and $m_x\sim m_y$.

Unlike the usual quadratic dependence expected in a Fermi liquid,
the frequency dependence of the imaginary part of the self-energy
in a nested region of the Fermi surface, or at Van Hove
singularities is known to be linear:

\begin{equation} {\rm Im} \Sigma_2 ( \bk , \ek ) \propto | \ek |
\label{hotspot}
\end{equation}
At the FS parts away from the hot spots, the leading contribution
to the two loop self-energy, when the Fermi surface is near a Van
Hove singularity, comes from diagrams where the polarizability
bubble, $\susc$ expressed in Eq.(\ref{polarizability}), involves
transitions near the saddle point\cite{HR95}.

Only particles in the vicinity of hot spots on the FS are strongly
scattered and present an anomalously large life-time, while away
from the hot spots the single particle lifetime follows the
Landau's Fermi liquid theory energy dependency. At some values of
the band filling the FS is near a nesting situation, as shown in
Fig.[\ref{FS-Desnudas}], then  the polarizability at low momenta
is similar to that of a one dimensional Fermi liquid, due to the
flat FS regions. The susceptibilities can be written as:
\begin{equation} \susc \sim \left\{ \begin{array}{lr} W^{-1}
\tilde{\Pi}_{\rm vH}
\left( \frac{\omega}{ m^* | \bq |^2} \right) &{\rm Van \, \, Hove} \\
W^{-1} \tilde{\Pi}_{\rm 1D} \left( \frac{\omega}{v_{\rm F} | \bq
|} \right)&{\rm nesting}
\end{array} \right. \label{susc} \end{equation} where $m^*$ is an average of the
second derivative of the bands at the saddle point. Note that, in
both cases, the density of states is proportional to the inverse
bare bandwidth $W^{-1} \sim t^{-1} , t'^{-1}$.

The imaginary part of the second order self-energy near the
regular regions of the Fermi surface can be written as\cite{HR95}:

\begin{equation}
{\rm Im} \Sigma_2 ( \bk , \ek ) \sim \int_0^{\ek} d \omega
\int_0^{q_{\rm max}} d q \, {\rm Im} \Pi ( q , \omega ) \label{HR}
\end{equation}

where $q_{\rm max} \sim | \Lambda | / v_{\rm F}$, and $v_{\rm F}$
is the Fermi velocity in these regions. By combining
Equations(\ref{susc}) and (\ref {HR}), we find:

\begin{equation}
{\rm Im} \Sigma_2 ( \bk , \ek ) \propto \left\{ \begin{array}{lr}
\ek^{3/2} &{\rm Van \, \, Hove} \\ \ek^2 &{\rm nesting}
\end{array} \right. \label{susc_reg}
\end{equation}

According to Eq.(\ref{susc_reg})  the usual Fermi liquid result is
recovered for the regular parts of the Fermi surface near almost
nested regions. This result arises from the fact that the small
momentum response of a quasi--one--dimensional metal does not
differ qualitatively from that predicted by Landau's theory of a
Fermi liquid while, close to  the Van Hove singularities, the
energy dependence of the ${\rm Im} \Sigma_2 ( \bk , \ek )$
presents anomalous exponents.

The effects induced by  inflexion points  have been addressed in
\cite{GGV97b,FG02}, where the instabilities of anisotropic 2D
systems are analyzed. Near the inflexion points, the dispersion
relation can be expanded about the Fermi level and satisfies

\begin{equation}
\ek \approx \left\{\begin{array}{cc}
  v_Fk_{\parallel}+b_1 k_{\perp}^3 & \textrm{inflexion point} \\
  v_Fk_{\parallel}+b_2 k_{\perp}^4 & \textrm{special inflexion point} \\
\end{array}\right.
\end{equation}

where $k_{\parallel}$ is the momentum parallel to the FS relative
to $k_F$, $k_{\parallel}=(\bk-\bk_F)_{\parallel}$, $k_{\perp}$ is
the momentum perpendicular to the FS relative to $k_F$,
$k_{\perp}=(\bk-\bk_F)_{\perp}$, and $b_{1,2}$ are constants. The
special inflexion points lie along a reflection symmetry axis of
the BZ\cite{FG02} (the $k_x=k_y=k_0$ points). We use the
techniques previously developed in Ref.[\cite{GGV97b,FG02}] to
obtain the second order self-energy near an inflexion point:
\begin{equation}
{\rm Im} \Sigma_2 ( \bk , \ek ) \propto \ek^{3/2}
\end{equation}

In the case of an special inflexion point, where the Fermi surface
changes from convex to concave and a pair of inflexion points are
generated for $\ef = \epsilon_0$ and ${\bf \vec{k}} \equiv ( k_0 ,
k_0 )$ defined earlier, the imaginary part of the self-energy
behaves as ${\rm Im} \Sigma_2 ( \bk , \ek ) \propto \ek^{5/4}$.

Once the imaginary part of the self-energy is known, we can obtain
the real part of the self-energy from it by means of a Kramers
Kronig transformation. Although ${\rm Im} \Sigma_2 ( \bk , \ek )$
has been given for $\omega= \ek$, since we are in the weak
coupling regime $U \simeq \Lambda \ll \epsilon_{\rm F}$, where
$\epsilon_{\rm F}$ is of the order of the non-interacting
bandwidth, then the imaginary part of the self-energy associated
to a state with energy $\epsilon_{{\bf \vec{k}}}$ is only
significant in an energy range $- \Lambda  + \epsilon_{{\bf
\vec{k}}} \le
         \omega \le \Lambda + \epsilon_{{\bf \vec{k}}}$.
We assume that one can approximate ${\rm Im} \Sigma_{\bf
{\vec{k}}} ( \omega )$ in this range by an expansion on $( \omega
- \epsilon_{{\bf \vec{k}}} ) / W$, where $W$ is an energy scale of
the order of the bandwidth in the non interacting problem, and
keep only the lowest order term. This approximation neglects
contributions from a region of energies centered around $| \omega
- \epsilon_{{\bf \vec{k}}} | \simeq \Lambda$ and of width $\delta
\Lambda$, which is, at most, a fraction of $\Lambda$. The
contribution of the Kramers Kronig transformation performed in
this region of energies is, at most, of order ${\rm Im}
\Sigma_{\bf \vec{k}} ( \Lambda  )$ and does not modify the
dependence of ${\rm Re} \Sigma_{\bf \vec{k}} ( \epsilon_{\bf
\vec{k}} )$ on the local properties of the Fermi surface.
Therefore we can obtain the real part of the self-energy from the
imaginary part by a Kramers-Kronig transformation, and restricting
the frequency integral to the interval $0 \le \omega \le \Lambda$,
we obtain:
\begin{equation}
{\rm Re} \Sigma_2 ( \bk , \ek ) \propto - g^2 | \Lambda | \times
\left\{
\begin{array}{lr} \log^2 \left( \frac{\Lambda}{\ek} \right) &{\rm Van \, \, Hove}
\\ \log \left( \frac{\Lambda}{\ek} \right) &{\rm nesting} \end{array}
\right. \label{real_hot}
\end{equation} where the negative sign is due to the fact that
it is a second order contribution in perturbation theory, and $g$
is a dimensionless coupling constant of order $U/W$. The sign is
independent of the sign of $U$ in Eq.(\ref{hamil}). In the regular
parts of the Fermi surface, Eq.(\ref{susc_reg}) leads to:
\begin{equation}
{\rm Re} \Sigma_2(\bk,\ek) \propto \left\{ \begin{array}{lr} - g^2
\frac{| \Lambda |^{3/2}}{W^{1/2}} &{\rm Van \, \, Hove} \\ - g^2
\frac{| \Lambda |^2}{W} &{\rm nesting}
\end{array} \right. \label{real_reg} \end{equation} where the
additional powers in $W$ arise from the $m^*$ and $v_{\rm F}$
factors in the susceptibility, expressed in  Eq.(\ref{susc}).

In the limit $\Lambda / W \rightarrow 0$, the different dependence
on $\Lambda$ of the self-energy corrections at different regions
of the Fermi surface is enough to give a qualitative description
of the changes of the Fermi surface. For instance, when the non
interacting Fermi surface is close to the saddle point, ${\bf
\vec{k}} \equiv a^{-1} ( \pm \pi , 0 ) , a^{-1} ( 0 , \pm \pi )$,
the self-energy correction is negative and highest in this region.
Note that the logarithmic divergences in Eq.(\ref{real_hot}) are
regularized by the temperature or elastic scattering.

At band fillings where the FS lies close to the Van Hove
singularities, most part of the low energy states close to the
Fermi energy, are around the saddle points $(0,\pm\pi)$ and
$(\pm\pi,0)$ (see Fig.(\ref{FS-Desnudas})). Strong screening
processes arise due to the big density of states at these points,
and if the chemical potential of the system is kept fixed, i.e.,
the system is in contact with a charge reservoir, the number of
particles varies and the Fermi energy tends to be pinned at the
Van Hove singularities \cite{G01}. Then,in order to remove the
Fermi surface from a Van Hove point or nesting situation, a large
number of electrons must be added to the regular regions. When the
points of the FS near these hot spots are at distance $k$ from the
hot spot, the change in the self-energy needed to shift the Fermi
surface by an amount $\delta k$ is, using Eq.(\ref{real_hot}),
\begin{equation}
\delta \Sigma \propto g^2 \Lambda \frac{\delta k}{k}
\label{shift_hot}
\end{equation}
with additional logarithmic corrections near a Van Hove
singularity. Near the regular regions of the Fermi surface, a
shift in energy of order $\delta \Sigma$ leads to a change in the
momentum normal to the Fermi surface of magnitude $\delta k_{\rm
reg} \sim \delta \Sigma / v_{\rm F}$. The area covered in this
shift gives the number of electrons which are added to the system
near the regular regions of the Fermi surface. We find
\begin{equation} \delta n \sim k_{\rm max} \delta k_{\rm reg} \sim
g^2 \frac{k_{\rm max} \Lambda}{v_{\rm F}}  \frac{\delta k}{k}
\label{shift_n} \end{equation} where $k_{\rm max} \sim a^{-1}$
determines the size of the regular regions of the Fermi surface.
The value of $\delta n$ diverges as the Fermi surface moves
towards the hot spot, $k \rightarrow 0$. Hence, the number of
electrons needed to shift the FS away from the hot spot also
diverges. This result has been obtained from calculations at fixed
chemical potential[\cite{GGV96,G01}], where the presence of a
charge reservoir is considered, with regular self-energy
corrections. This situation has particular interest when studying
the physics of high-$T_c$ cuprates, where doping of the CuO$_2$
layers and interactions with the rest of the perovskite structure
are important. The pinning of the Fermi level to the Van Hove
singularity has been investigated in the 2D $t-t'$ Hubbard model
by RG techniques\cite{IKK02}, taking into account the formation of
flat bands due to the renormalization of the electron spectrum.
The pinning of the Fermi level to the Van Hove singularities is
found without making use of a reservoir, and the chemical
potential of the system remains practically constant in a range of
dopings near the Van Hove filling.

\subsection{Self-energy corrections to regular Fermi surfaces.}
In this section we study a 2D system at a band filling which
yields a curved FS, slightly anisotropic, in the absence of
singularities. Near the Fermi surface, by choosing an appropriate
coordinate system, the electronic dispersion can be approximated
by:
\begin{equation}\label{epsilon-k}
\ek = v_{F}k_{\perp}+\beta k_{\parallel}^{2} \label{dispersion_2}
\end{equation}

where $\ek=\varepsilon (\bk)-\ef$, $k_{\perp}$ is the momentum
perpendicular to the FS relative to $k_F$,
$k_{\perp}=(\bk-\bk_F)_{\perp}$, $k_{\parallel}$ is the momentum
parallel to the FS relative to $k_F$,
$k_{\parallel}=(\bk-\bk_F)_{\parallel}$, $\vf$ is the Fermi
velocity at any particular point
$\vf=\hat{\mathbf{n}}_{\perp}\cdot\mathbf{\nabla}\varepsilon(\bk)$,
and $\beta$ is related to the local curvature of the Fermi surface
$b=\hat{\mathbf{n}}_{\parallel} \cdot
\left(\mathbf{\nabla}^2\varepsilon(\bk)\right)
\hat{\mathbf{n}}_{\parallel}$, by $\beta=bv_F/2$. This expansion
implies, without assuming any rotational symmetry, a FS locally
indistinguishable from a circular one, where the energy
Eq.(\ref{epsilon-k}) would correspond to a radius $k_F=m_Fv_F$,
where we have renamed $\beta=1/2m_F$, being $m_F$ the effective
mass. The Fermi velocity $v_F$ and the FS curvature $b$, are
functions of $t , t', \ef$ and the position along the Fermi line.
We calculate the second order diagram of Fig.(\ref{selfenergy}),
assuming that the main contribution to the self-energy arises from
processes where the momentum transfer is small, forward scattering
channel, or from processes which involve scattering from the
region under consideration to the opposite part of the Fermi
surface, {\it i. e.}, backward scattering (in Appendix A we give
the expressions of the polarizability for these two channels).
This assumption can be justified by noting that we are considering
a local Hubbard interaction, which is momentum independent, so
that the leading effects are associated to the structure of the
density of states. The processes discussed here are those
joining the regions which have the highest density of states.

From Eq.(\ref{selfenergy-full}) we can obtain the imaginary part
of the self-energy, which describes the decay of quasiparticles in
the region under consideration and that it is independent of the
cutoff $\Lambda$. The contribution from forward scattering
processes is:
\begin{equation}\label{Im-Sigma-Forw}
\rm {Im}\Sigma_2(\bk,\omega)=\frac{3}{64}\frac{U^2
a^4}{\sqrt{2}\pi^2}\frac{\omega^2}{\vf^2|\beta|} \ .
\end{equation}


The quadratic dependence of energy is expected, and consistent
with Landau's theory of a Fermi liquid. This contribution diverges
as $\vf \rightarrow 0$, that is, when the Fermi surface approaches
a Van Hove singularity, or as $| b | \rightarrow 0$ which signals
the presence of an inflexion point or nesting. The contribution
due to backward scattering is exactly the same as that from
forward scattering, Eq.(\ref{Im-Sigma-Forw}), with the same
numerical prefactors.

Using a Kramers-Kronig transformation, and integrating in the
interval $0 \le \omega \le \Lambda$, we obtain:

\begin{equation}
\rm
Re\Sigma_2\!(\bk,\omega)\!=\!-\frac{3}{64}\frac{U^2a^4}{\sqrt{2}\pi^3}\frac{1}{v_{\rm
F}^2|\beta|}
\!\!\left[\!\Lambda^2\!+\!2\Lambda\omega\!+\!2\omega^2\!\log\!\left|\frac{\Lambda-\omega}{\omega}\right|\right]
\label{resigma}
\end{equation}

From the experimental point of view the determination of the
scattering rate (${\rm Im} \Sigma(\bk,\omega)$) presents
particular interest and  big effort has been devoted in order to
obtain it: by ARPES because of the momentum and energy resolved
measurements \cite{Valla00, Kami05, Kordy05,Ingle05} and recently
by electrical transport experiments at microwave
frequencies\cite{Ozcan06}.

\begin{figure}
\resizebox{8.2cm}{!}{\includegraphics[]{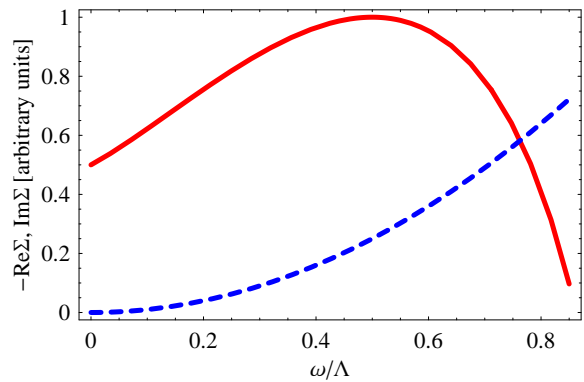}}
\caption{(Color online) Real and imaginary parts of the self
energy as a function of the energy $\omega$: full red line $-{\rm
Re} \Sigma$, dashed blue line ${\rm Im} \Sigma$}\label{Sigma-w}
\end{figure}

The extraction of the correlation functions from the experimental
data is a complicated task and, although many theoretical
approximations exist, the computation of correlation effects is
also difficult. From ARPES results in  underdoped and optimally
doped cuprate samples \cite{Kami05} an anisotropic scattering rate
around the Fermi surface have been found and the bare Fermi
velocity has been directly obtained. By using a different
methodology the real and imaginary parts of the self-energy has
been obtained from photoemission data by a self-consistent
procedure \cite{Kordy05}.

In Fig.[\ref{Sigma-w}] we represent the self-energy as a function
of the frequency according to results from
Eqs.(\ref{Im-Sigma-Forw},\ref{resigma}), where the linear
(quadratic) behavior of the real (imaginary) part of the
self-energy at low frequencies, typical of a Fermi liquid system,
is recovered. We find a qualitative agreement with the low-energy
part of the self-energy functions extracted self-consistently from
the experiments in Ref.[\cite{Kordy05,X06}]. It should be notice
that we consider here the electron-electron scattering only. The
impurity and electron-phonon scattering will, no doubt, cause
finite life-time and energy renormalization of the excitations but
our main concern is the self-energy due to electro-electron
correlation. The impurity scattering term can be considered to be
isotropic (from an isotropic distribution of static impurity
scatterers) an it will give a constant term in ${\rm Im}
\Sigma(\bk,\omega)$. The electron-phonon self-energy can be
assumed to be small at low temperature. Then, the assumption that
the dominant scattering mechanism is the electron-electron
interaction in the systems under study is not
unreasonable\cite{Ingle05}. The essential features of the
electron-phonon coupling are explained in Ref.[\cite{ZCDNS06}]
both in the superconducting and normal states. An effective $\rm
Re\Sigma$ is extracted from ARPES measurements and by fitting and
modelling of the data information about bosonic excitations.

As a last result, from the real part of the self-energy
Eq.(\ref{resigma}), we can calculate the quasiparticle-weight

\begin{equation}
 Z_{\bk_{F}}=\left(1-\frac{\partial
\rm{Re}\Sigma(\bk_F,\omega)}{\partial
\omega}\right)^{-1}_{\omega\rightarrow 0}
\end{equation}

which for our case reads

\begin{equation}\label{Z}
 Z_{\bk_{\rm
F}}=\frac{1}{1+\frac{3\sqrt{2}a^4}{64\pi^3}\frac{U^2\Lambda
}{|\beta|\vf^2}}
\end{equation}


In the weak coupling regime, the quasiparticle-weight would be
minimum either if the Fermi velocity becomes very small (Van Hove
singularity, consistent with results in Ref.[\cite{Kata04}]) or if
the curvature of the Fermi surface changes sign (inflexion point).

Finally, from Eq.(\ref{resigma}) we obtain the expression which
gives the zero frequency limit of the real part of the self-energy
\begin{equation}\label{resigma0}
\rm {Re}\Sigma_2(\bk,\omega=0)=-\frac{3}{64}\frac{U^2
a^4}{\sqrt{2}\pi^3}\frac{\Lambda^2}{v_{\rm F}^2|\beta|}
\end{equation}
which renormalizes the FS according to Eq.(\ref{dFS}).

\subsection{Application to Fermi surfaces of copper oxides superconductors}

As stated in the Introduction, copper oxides superconductors have
a strongly anisotropic layered structure. Cuprates present
anomalous properties in many physical aspects and are one of the
main challenges to condensed matter physics. The peculiarities of
the phase diagram have been addressed by different techniques and
no consensus has been reached. Experimental results indicate a
behavior far from the Fermi liquid, and the changes induced by
doping on the ground state character of the normal state add
complexity to the problem. Both, the low dimensionality ($CuO_2$
planes) together with the strong electronic correlations, have to
be taken into account to understand the low energy excitation
spectrum of the cuprates. The effects of the strong correlations
on the Fermi surface shapes of the cuprates is a hot issue.
Recently Civelli \emph{et al} \cite{Kotl05} have addressed the
problem by using an extension of the dynamical mean-field theory,
the cellular dynamical mean-field theory (CDMFT) which allows the
study of $k-$dependent properties. They study the 2D Hubbard model
in the strongly correlated regime ($U=16t$). A strong
renormalization of the FS shape, due to interactions, is found
together with a momentum space differentiation: appearance of hot
and cold regions in the Brillouin zone.

\begin{figure}
\resizebox{8.5cm}{!}{\includegraphics[]{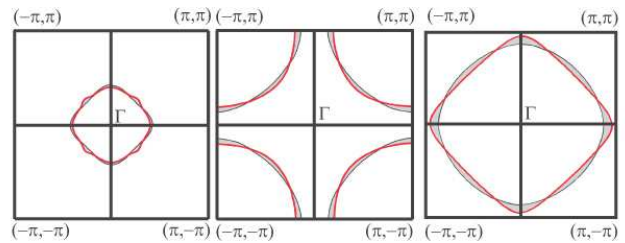}}
\caption{(Color online) Deformations induced by the interactions
on the FS of the $t-t'$ Hubbard model. The axes are labelled in
units of $a^{-1}$, where $a$ is the lattice constant. Black line
represents the unperturbed FS while the red line represents the FS
corrected by the interaction. For $t'/t=-0.3$ (a): high doping
range and (b) close to half filling. For $t^{\prime}/t=+0.3$ (c)
close to half filling.} \label{holeFS}
\end{figure}

\begin{figure}[hb]
  \includegraphics[width=8.5cm]{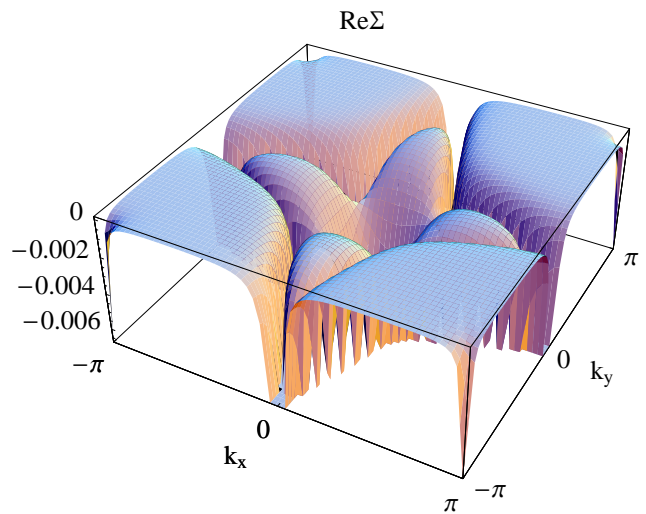}\\
  \caption{{\label{Sigma-3D-hole-cuprates}}(Color online) Real part of the self-energy for
  $t^{\prime}=-0.3t$ represented in the square Brilluoin zone.}
\end{figure}

We study as well the two dimensional Hubbard model on a square
lattice, and we will consider hopping amplitudes $t_{ij}$ to
nearest-neighbors $t$ and to next nearest neighbors $t'$. Since we
are in the perturbative regime we consider local weak coupling
instead of the strong coupling addressed in reference
\cite{Kotl05}. We adopt the hopping values $t=-1$ and $t'=-0.3t$
which mimic the hole-doped cuprates ($t/t' < 0$), and we will
consider two different doping levels. In Eq.(\ref{resigma0}) we
can see how the self-energy corrections due to electron-electron
interactions depend on local features of the non-interacting FS,
as the Fermi velocity $\vf$ and the curvature $b$. For the
dispersion relation given by Eq.(\ref{dispersion}),  the
expressions derived for the Fermi velocity and the curvature
(taking for simplicity $a=1$) at the Fermi level, are:

\begin{widetext}
\begin{equation}
v_F(\bk)=\sqrt{(t+2t'\cos{k_y})^2\sin^2\!k_x+(t+2t'\cos{k_x})^2\sin^2\!k_y}
\end{equation}

\begin{equation}\label{curvature}
b(\bk)=-\frac{(t+2t'\cos{k_x})(t+2t'\cos{k_y})[t\cos{k_y}\sin^2\!k_x+2t'\cos^2\!k_y\sin^2\!k_x+(t\cos{k_x}-t'(-3+\cos{2k_x}))\sin^2\!k_y]}{\left[(t+2t'\cos{k_y})^2\sin^2\!k_x+(t+2t'\cos{k_x})^2\sin^2\!k_y\right]^{3/2}}
\end{equation}
\end{widetext}

where $\bk$ stands for the Fermi momentum $\bk_F$. These
expressions illustrate the momentum dependence of the self-enegy
corrections. At high doping, the FS presents an almost square
shape with rounded corners. We find that the self-energy
corrections are stronger for the regions with the smallest
curvature at the diagonal parts of the BZ. This behavior, shown in
Figure[\ref{holeFS}](a) for $\ef = -2.3 \approx 8t^{\prime}$, is
similar to the result obtained by Freire \emph{et al} in
\cite{FCF04} for the renormalization of a flat FS by a two-loop
field theory RG approach, where interactions induce a small
curvature to the bare flat FS. They found as well that, the
renormalized FS, becomes truncated due to the interactions, not
found here. Next we will consider a lower doping level. By
changing the filling, the FS shape varies, and close to
half-filling, for $\ef =-0.9=3t^{\prime}$, the FS has the form
shown in Figure[\ref{holeFS}](b). The change in shape
qualitatively agrees with the doping evolution of $k_F$ measured
by ARPES on cuprates \cite{Yosh,Kami}, and the FS shape is similar
to the FS reported in different experiments. At this doping, close
to half-filling, the self-energy corrections enhance the hole-like
curvature around $(\pm\pi,\pm \pi)$ and $(\pm\pi,\mp \pi)$, and
flatten the FS close to the $(\pm\pi, 0)$ and $(0,\pm \pi)$ points
of the BZ as it is shown in Fig.[\ref{holeFS}](b). This result
coincides with the renormalization found in \cite{Kotl05} even
though they are in the strongly correlated regime.

Corrections found in Figure[\ref{holeFS}] can be understood by
looking at Figure[\ref{Sigma-3D-hole-cuprates}], where the real
part of the self-energy is depicted at $\omega=0$ in the square
BZ. At the central region of the BZ near the $\Gamma$ point, the
main corrections occur around the diagonal of the BZ, as we have
found in the high doping case, where the flat parts of the
non-interacting FS become curved. At this point, the larger
contribution to ${\rm Re} \Sigma(\bk,\omega)$ will be due to the
curvature of the Fermi surface, which is almost flat in the nodal
region at this value of the band filling. In fact, it can be
observed in Figure[\ref{Sigma-3D-hole-cuprates}], following the
$(-\pi,-\pi)-(\pi,\pi)$ diagonal line, that the minima of $\rm Re
\Sigma(\bk,\omega)$ coincide with the minima of the curvature,
that corresponds to the maximum correction to the non-interacting
FS.

Looking again at Figure[\ref{Sigma-3D-hole-cuprates}] we see that,
for higher fillings, the corrections are more pronounced where
inflexion points start to appear in the Fermi surface (see
Figure[\ref{FS-Desnudas}]). The first of these inflexion points
$k_0$ occurs in the diagonal of the BZ, in the nodal direction.
Once this first inflexion point appears, if we increase the
occupation towards half-filling, new inflexion points merge in
each FS and they distribute themselves symmetrically with respect
to the diagonals of the BZ. This is why the \emph{divergence
valley} has this star-like structure. Finally, at lower doping
levels, as that represented on the right panel of
Figure[\ref{holeFS}], the Fermi line reaches the region closer to
the border of the BZ, and the main corrections appear at the
proximity of the antinodal points $(0,\pm\pi)$, $(\pm\pi,0)$. As
can be observed in Figure[\ref{Sigma-3D-hole-cuprates}], ${\rm Re}
\Sigma(\bk,0)$ shows pronounced dips close to the saddle points,
and therefore the FS is renormalized in this region. These minima
are due to Van Hove singularities where the $\vf$ vanishes.


This result agrees with one-loop functional RG calculation of the
self-energy in the weak coupling regime of the 2D $t-t'$- Hubbard
model at Van Hove band fillings \cite{Kata04} where vanishing of
the quasiparticle-weight on approaching the antinodal points is
found. Away from the Van Hove fillings a quasiparticle peak, with
small spectral weight, emerges at $(\pi,0)$ and $(0, \pi)$.

The case of an electron doped system can be analyzed by a
particle-hole transformation of the Hamiltonian which reverses the
sign of $t'$, ($t/t' > 0$). For $t'=+0.3t$ we find that, close to
half-filling, the self-energy corrections are stronger in the
proximity of the saddle points, where $\vf \rightarrow 0$. As can
be seen in Fig.[\ref{holeFS}](c), the corrected FS is closer to a
nesting situation than the bare FS.

Our results, near half-filling,  are in overall agreement with
those of reference[\cite{Kotl05}], although we find that the
self-energy corrections are stronger at the antinodal region in
both, hole-like and electron-like, Fermi surfaces.

\subsection{Application to Fermi surface of Sr$_2$RuO$_4$}

\begin{figure}[htb]
  \includegraphics[width=9cm]{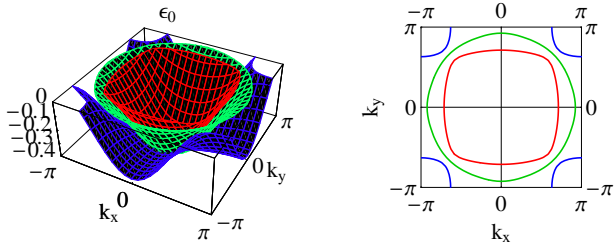}\\
  \caption{{\label{Bands+FS-SrRuO}}(Color online) Left: Band structure of Sr$_2$RuO$_4$ obtained from Eq.(\ref{dispersion-SrRuO})
  using the parameter values given in Ref.\cite{CM06}. The blue sheet corresponds to the $\alpha$-band,
  the red sheet is the $\beta$-band and the green sheet correspond to the $\gamma$-band. Right: Corresponding Fermi surface.}
\end{figure}

Sr$_2$RuO$_4$ is a highly anisotropic layered compound, with an
electrical anisotropy of about 4000 \cite{Mackenzie03}. It has a
strongly two-dimensional electronic structure and exhibits a good
Fermi liquid behavior below 30K, as probed by bulk transport
measurements\cite{Mackenzie03}. The Fermi surface measured by
ARPES matches the de Haas van Alphen measurements and it can be
well described by band structure calculations. Therefore
Sr$_2$RuO$_4$ is considered a correlated 2D material. As it occurs
in the cuprates, the competition between superconducting and
magnetic instabilities play an important role in the low energy
physics of Sr$_2$RuO$_4$ which, with a critical temperature of
about $T_c \approx 1.5 \rm K$, presents an unconventional
superconductivity with a $p$-wave and spin-triplet
pairing\cite{RS95}. This material has three relevant
bands\cite{Berge03C}, $\alpha$, $\beta$, and $\gamma$, of $t_{2g}$
symmetry, formed from the $4d$-orbitals of the Ru$^{4+}$ ion. The
$\{\alpha,\beta\}$ bands are derived from $\{d_{xz},d_{yz}\}$
orbitals and form two quasi-one-dimensional bands along the
directions $z$ and $y$ respectively, that are weakly hybridized.
The $\gamma$ band is derived from the $d_{xy}$-orbital and
disperses into a real 2D band. In Figure[\ref{Bands+FS-SrRuO}] are
depicted the electronic structure, left panel and corresponding
FS, right panel, calculated following the dispersion relation

\begin{widetext}
\begin{eqnarray}\label{dispersion-SrRuO}
\varepsilon_{\gamma}(\bk)&=&-2t_{1\gamma}(\cx+\cy)-4t_{2\gamma}\cx\cy-\varepsilon_{0\gamma}\nonumber\\
\varepsilon_{\alpha,\beta}(\bk)&=&-(t_{1\alpha,\beta}+t_{2\alpha,\beta})(\cx+\cy)\nonumber\\
&&\pm\sqrt{((t_{1\alpha,\beta}-t_{2\alpha,\beta})(\cx-\cy))^2+16t_{3\alpha,\beta}^2\sx^2\sy^2}-\varepsilon_{0\alpha,\beta}
\end{eqnarray}
\end{widetext}

\begin{figure*}[htb]
  \includegraphics[width=18cm]{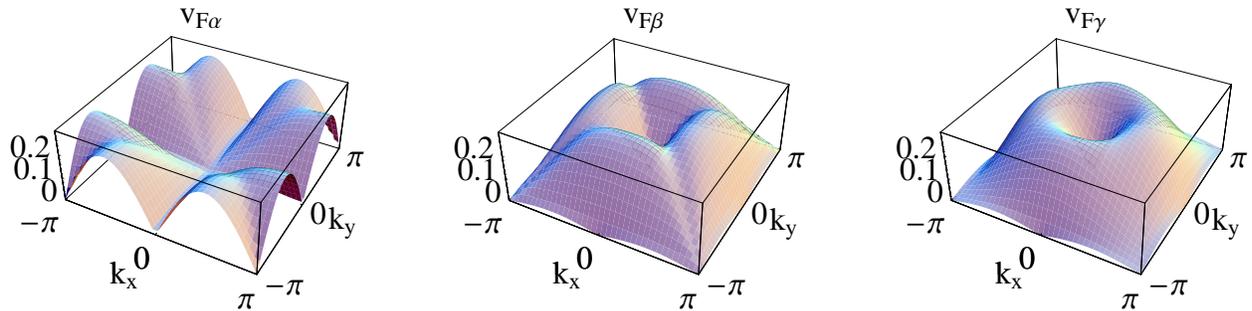}\\
  \caption{{\label{vF-3D}}(Color online) Fermi velocity of the non-interacting bands: Left graph corresponds to the $\alpha$-band,
  center graph corresponds to $\beta$-band and right graph corresponds to $\gamma$-band of Sr$_2$RuO$_4$.}
\end{figure*}

The tight-binding parameters are taken from
references[\cite{Berge03C,CM06}]. This FS is in good agreement
with that reported by ARPES experiments and obtained by band
calculations. As before the self-energy corrections to the three
sheets of the FS depend on the Fermi velocity and the curvature
(not given here due to its size). In Figure[\ref{vF-3D}] we can
see the momentum dependence of the Fermi velocity for the three
bands. It is easy to appreciate that the minima of the three plots
are, besides in the $\Gamma$ point, in the antinodal points,
$(\pm\pi,0)$ and $(0,\pm\pi)$, and in the diagonal of the BZ,
$(\pm\pi,\pm\pi)$ and $(\pm\pi,\mp\pi)$.

\begin{figure*}[htb]
  \includegraphics[width=18cm]{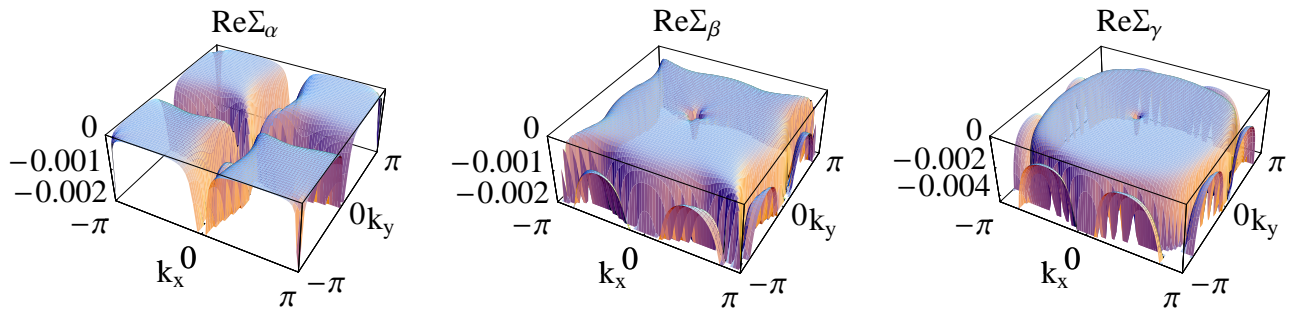}\\
  \caption{{\label{Sigma-3D-SrRuO}}(Color online) Real part of the self-energy corrections for the three bands of Sr$_2$RuO$_4$ in the first
  square BZ: Left graph corresponds to the
  $\alpha$-band, middle to the $\beta$-band, and right graph corresponds to $\gamma$-band.
  We have used the parameter values $U=0.01$ and $\Lambda=1$. Notice the different scales in the three graphs.}
\end{figure*}

The momentum dependence of the real part of the self-energy for
the three bands is shown in Figure[\ref{Sigma-3D-SrRuO}]. In the
left hand side we can see the correction to the $\alpha$-sheet,
that is significant only either when the Fermi line lies near the
$\Gamma$-point or in the proximity of the corners of the BZ,
$(\pm\pi,\pm\pi)$ points. In a recent experimental
work\cite{Ingle05}, the form of $Im\Sigma(\omega)$ extracted for
the bulk $\alpha$-band from the ARPES spectra, is found to be
consistent with a Fermi liquid, and that the quasiparticles
residing in the surface layer $\alpha$-band, show similar
many-body interactions. (Notice that the analysis and fitting
procedure over raw data of this band made in Ref.[\cite{Ingle05}]
leads to self-energy curves of the form of Fig.[\ref{Sigma-w}]).
In the central figure, the contribution corresponding to the
$\beta$-band is shown. Here we can see that the correction is
maximum (most negative) in the zones between the diagonals, due to
the nearly flat regions of the Fermi line corresponding to the
$\beta$-band.

Finally, the corrections to the $\gamma$-band are shown in the
right hand side of Figure[\ref{Sigma-3D-SrRuO}] (notice the
different scales in the three graphs). In this case the
corrections are maxima near the antinodal points $(\pm\pi,0)$ and
$(0,\pm\pi)$, due to the proximity of this band to a Van Hove
point. In reference\cite{CM06} it is pointed out that calculated
$\gamma$-band properties depend very sensitively on how close it
approaches the Van Hove points ($\pi$,0), (0,$\pi$).

\begin{figure}[htb]
  \includegraphics[width=7cm]{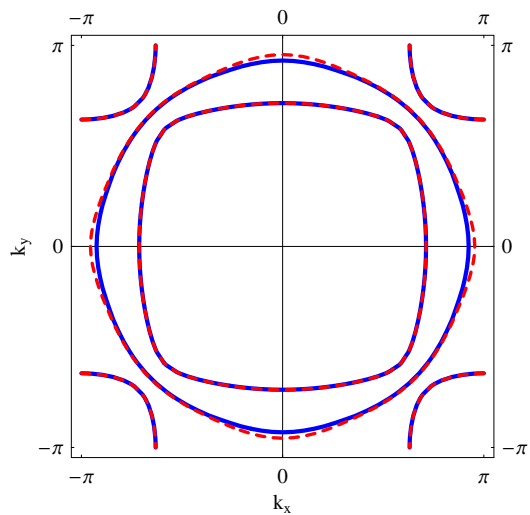}\\
  \caption{{\label{DressedFS-SrRuO}}(Color online) Bare (continue blue) and interacting (dashed red) Fermi surface of Sr$_2$RuO$_4$.}
\end{figure}

The main corrections occur at the $\gamma$-band as it is shown in
Figure[\ref{DressedFS-SrRuO}] where the bare and renormalized FS
are depicted. Here non-interacting FS (full blue line) is changed,
due to electron-electron correlations, to the interacting FS
(dashed red line). This result agrees with photoemission
measurements which indicate that the $\gamma$-band has much
stronger interactions and plays a dominant role at low
temperature\cite{KVF05}.

The corrections in the $\alpha$ and $\beta$ bands are due to
curvature effects, while the corrections to the $\gamma$ band are
due to Fermi velocity effects, because of the proximity of this
band to the saddle points, as it has been pointed out above. Then,
in agreement with ARPES measurements\cite{Ingle05} and other
theoretical results\cite{CM06}, the main corrections induced in
the FS of Sr$_2$RuO$_4$ occur at the $\gamma$-band corresponding
sheet. The importance of the FS geometry has been analyzed in
Ref.[\cite{CM06}] where the non-analytic corrections to the
specific heat and susceptibility of a 2D Fermi liquid have been
considered and the results applied to Sr$_2$RuO$_4$. Both, the
dependence of the $\gamma$ band properties on how close the band
approach the Van Hove points and the dominant interaction in the
$\gamma$ band are found in Ref.[\cite{CM06}], because it has the
highest density of states at the Fermi level, as well as the
largest mass and susceptibility enhancements\cite{Berge03C}.

Similarly, the importance of the band structure properties of
these materials can be seen in the context of multi-layer
ruthenates, for which the proximity of their Fermi surface to a
Van Hove singularity can give rise to a quantum critical end-point
in the magnetic phase diagram, as it is found, within a mean-field
analysis, in Ref.[\cite{BS04}].

\section{Conclusions.}

The unconventional physics shown by anisotropic materials, as
high-T$_c$ superconductors and ruthenates, poses big difficulties
when a theoretical model has to be chosen. On the other hand,
although it is generally accepted that the many-body interactions
of electrons play a key role in the underlying physics of these
compounds and may be related to the occurrence of
superconductivity in the cuprates, a consensus has not been
reached about the origin of important features widely observed by
different experimental techniques. The anisotropy momentum space
shown by many electronic properties of the planes adds complexity
to the possible theories.

In the cuprates, the pseudogap phase, metallic but with a broken
Fermi surface (segments known as Fermi arcs\cite{KNR06}) is an
example of the remarkable momentum dependence of the interactions.
Furthermore the band renormalization observed by ARPES in
different families of high T$_c$ superconductors, known as kink in
the dispersion, which indicates a strong coupling to a collective
mode (see Ref.[\cite{ZCDNS06}] and references therein), shows
different energy scales and temperature dependence at the nodal
and antinodal regions of the BZ, suggesting two different kinks of
different origin, which nature is under debate: both phonon and
magnetic mode have been suggested as possible causes.

More recently a high energy anomaly in the spectral function has
been observed in three different families of high-T$_c$
superconductors, and apparently in several ruthenate
compounds\cite{GGE06}. This anomaly indicates that the
quasi-particles at $\ef$ are dressed not only by the interactions
with bosons at low energy, but also by interactions at higher
energies. This peculiar high energy behavior together with the
unconventional low energy properties, poses a challenge on
theoretical models.

The knowledge of the dressed FS is crucial to understand the
behavior of correlated materials, especially when an effective
model is needed to explain the unconventional physics. We have
presented here a simplified way of taking into account the
self-energy corrections to the Fermi surface. We have made use of
the different dependence of the self-energy on the high energy
cutoff in order to analyze the main features of the changes of the
FS. The analysis presented here is valid only at weak coupling,
and we do not consider corrections to the interactions or to the
wave-function renormalization. On the other hand, the expressions
obtained are analytical and related to the local features of the
non-interacting FS in a simple way, so that they can be readily
used to get an estimate of the corrections expected.

The results suggest that the main self-energy corrections, which
are always negative in our scheme, peak when the FS is close to
the $( \pm \pi,0 ) , ( 0 , \pm \pi )$ points in the Brillouin
zone. If these contributions are cast as corrections to the
hopping elements of the initial hamiltonian, we find that the
nearest neighbor hopping, $t$,  is weakly changed (as it does not
contribute to the band dispersion in these regions). The next
nearest neighbor hopping, $t'$ which shifts the bands by $- 4 t'$
in this region, acquires a negative correction. This implies that
the absolute value of $t'$ grows when $t' > 0$, or decreases, when
$t' < 0$, in reasonable agreement with the results reported
in\cite{Kotl05}. Note that the tendency observed in our
calculation towards the formation of flat regions near these
points, when analyzed in higher order perturbation theory, will
lead to stronger corrections. Our results also confirm the pinning
of the FS near saddle points, due to the interactions. The
analysis presented here is consistent with  the measured Fermi
surfaces of the cuprates\cite{DHS03} and qualitatively agree with
the doping evolution reported by ARPES\cite{DHS03, Yosh, Kami}.
The self-energy corrections found for the FS of Sr$_2$RuO$_4$,
which mainly renormalize the $\gamma$ sheet, are as well in
qualitative agreement with ARPES measurements\cite{Ingle05} and
previous calculations\cite{CM06}. The broad spectrum of
experimental data available at this moment makes comparison
between results from different techniques one of the most
efficient methods to obtain information about response and
correlation functions of unconventional materials. To get an
estimation of the self-energy corrections to the Fermi surface in
a simple way, independent of the model, as the one here proposed,
is helpful in order to gain insight into many low-energy physics
aspects.

{\it Acknowledgments.} Funding from MEC (Spain) through grants
FIS2005-05478-C02-01 is acknowledged. We appreciate useful
conversations with A. H. Castro Neto, A. Ferraz, J. Gonz\'alez, G.
Kotliar, and M.A.H. Vozmediano.

\appendix
\section{polarizability}

In this Appendix the particle-hole polarizability for the forward
and backward scattering channels are given. These polarizabilities
are calculated in oreder to obtain the self-energy
Eq.(\ref{Im-Sigma-Forw}). The imaginary part of $\susc$ for the
forward channel is, using Eq.(\ref{polarizability}) and the
parametrization from Eq.(\ref{dispersion_2}):

\begin{equation}\label{Im-Gamma-Forw}
\rm
Im\Pi^F\!(\bq,\omega)=-\frac{U^2}{16\pi}\frac{|\omega|}{\beta\vf}\sqrt{\frac{2\beta}{\omega-M^F_{\bq}}}
\end{equation}

where $\bq$ is a small vector that connects two pieces close
together in the FS and $\rm M^F_{\bq}=v_Fq_\perp+\frac{2}{9}\beta
q_{\parallel}^2$. The argument of the square root has to be
positive, what gives an extra condition, $\rm \omega\,
sgn(\beta)>\rm M^F_{\bq}\, sign(\beta)$.

Similarly, for the backward scattering we obtain:

\begin{widetext}
\begin{equation}\label{Im-Gamma-Back}
\rm Im\Pi^B\!(\vec{\bf
Q}+\bq,\omega)=-\frac{U^2}{16\pi}\frac{1}{\beta\vf}\left\{\begin{array}{cc}
  \sqrt{\rm 2\beta(|\omega|+M^B_{\bq})}-\sqrt{\rm 2\beta(M^B_{\bq}-|\omega|)} & \textrm{if $|\omega|<\rm M^B_{\bq}\,sgn(\beta)$}\\
 \rm sgn(\beta)\sqrt{2\beta(|\omega|\rm sgn(\beta)+\rm M^B_{\bq})}&
\textrm{if\,\,\,\,\,\,$\rm |\omega|>|M^B_{\bq}|$}\\
\end{array}
\right.
\end{equation}
\end{widetext}

where $\bq$ is the deviation of the wave-vector from the vector
$\vec{\bf Q}$ that connects the region studied with the opposite
part of the FS, and $\rm M^B_{\bq}=\vf q_{\perp}-\frac{2}{9}\beta
q_{\parallel}^2$.

For $\bq=0$ we have that $\rm Im\Pi^B\!(\vec{\bf
Q}+\bq,\omega)\sim\sqrt{|\omega|}$, which agrees with the
results obtained in Ref.[\cite{KC06}], where  a spin-fluctuation model for
the $Q=2k_F$ instability is studied. For small $\omega$ and fixed $\bq$ we
have, expanding Eq.(\ref{Im-Gamma-Back}) up to first order in
$|\omega|/\rm M^B_{\bq}$ (for the case
$\rm|\omega|<M^B_{\bq}\,sgn(\beta)$), that $\rm Im\Pi^B\!(\vec{\bf
Q}+\bq,\omega)\sim|\omega|$, as expected for a
Fermi-liquid\cite{KC06}.

\bibliography{FSR}

\end{document}